\documentclass[12pt]{article}
\usepackage{amsmath, amssymb}
\usepackage{geometry}
\usepackage{setspace}
\usepackage{hyperref}
\usepackage{graphicx}
\usepackage{subcaption}

\begin{document}

\title{Equilibrium Analysis of Discrete Stochastic Population Models with Gamma Distribution}
\author{Haiyan Wang\\
School of Mathematical and Natural Sciences\\
Arizona State University\\
Phoenix, AZ 85069\\
haiyan.wang@asu.edu
}
\date{}
\maketitle

\begin{abstract}
This paper analyzes the stationary distributions of populations governed by the discrete stochastic logistic and Ricker difference equations at equilibrium examines with the gamma distribution. We identify mathematical relationships between the intrinsic growth rate in the stochastic equations and the parameters of the gamma distribution with a small stochastic perturbation. We present the biological significance of these relationships, emphasizing how the stochastic perturbation and shape parameter of the gamma distribution influence population dynamics at equilibrium. Furthermore, we identify two branches of the intrinsic growth rate, representing alternative stable states corresponding to higher and lower growth rates. This duality provides deeper insights into population stability and resilience under stochastic conditions.
\end{abstract}

\section{Introduction}
Modeling population dynamics has provided valuable insights into how populations grow, stabilize, or decline in response to environmental pressures, making it a fundamental aspect of ecology and biology. The logistic and Ricker difference equations are widely employed to describe population growth under resource constraints. These equations, in their deterministic forms, describe an initial phase of exponential population growth that slows as the population approaches the carrying capacity, representing the maximum number of individuals the environment can sustain. The dynamics of these deterministic models, including stable growth, periodic oscillations, chaotic behavior, or extinction, are strongly influenced by the intrinsic growth rate \cite{may1976simple,strogatz2018nonlinear,hilborn2000chaos}.

Real-world populations frequently experience random fluctuations driven by environmental variability, demographic stochasticity, and other unforeseen factors. These stochastic influences necessitate the use of stochastic equations to more accurately capture population dynamics under real-world conditions. By introducing randomness into population models, stochastic equations account for the variability and uncertainty inherent in natural systems.

Stochastic models offer distinct advantages over deterministic ones. While deterministic models predict fixed outcomes based on initial conditions, stochastic models describe the distributions of population sizes, providing a richer understanding of both the expected behavior and the range of potential fluctuations around it. This ability to account for randomness enables stochastic models to better represent the dynamics of real-world populations, particularly under fluctuating environmental conditions \cite{Allen, Renshaw,Matis2003,May2001,Dennis2016,Nasell2003,kot2001elements,Kemp1993}. The study of stochastic effects in logistic and Ricker models has been an active area of research, with a diverse body of literature exploring various extensions and applications of these models \cite{Braverman2013,Aktar2023,Yan2024}.

In this paper, we explore the steady states of modified stochastic logistic equation 
\begin{equation} \label{eq:logistic3}
X_{t+1} = r X_t \left( 1 - X_t^n \right) \epsilon_t,
\end{equation}
and the stochastic Ricker equation:
\begin{equation}
X_{t+1} = X_t e^{r(1 - X_t)} \epsilon_t
\label{ricker1}
\end{equation}
where positive integer \( n \geq 1 \) introduces a nonlinearity into the population growth term, \( X_t \) is a distribution of population size at time \( t \), \( r >0\) is the intrinsic growth rate,  $\epsilon_t$ is a small nonnegative perturbation distribution representing stochastic effects, assumed it is independent of $X_t$ and its mean is $1$ ($E[\epsilon_t] = 1$). We use the gamma distribution to analyze stochastic models at equilibrium. For general \( n \), we derive and analyze relationships of the intrinsic parameter $r$  with the gamma distribution parameters. Further we study the impact of the variance of $\epsilon_t$ on $r$ and therefore the dynamics of (\ref{eq:logistic3}) and (\ref{ricker1}).

Stochastic population models account for the inherent variability of natural systems by describing stationary distributions that capture fluctuations around a steady state. The gamma distribution has been widely recognized as a reliable approximation of stationary distributions in various ecological settings \cite{Dennis1984, Pielou1975, Engen1978}. The gamma distribution is particularly effective in studies of species like the \textit{Tribolium} beetle, where laboratory populations exhibit fluctuations around a mean equilibrium size driven by environmental variability \cite{Dennis1984, Peters1989, DennisCostantino1988, Costantino1981}. These findings highlight the gamma distribution's utility in modeling population dynamics under stochastic influences.

In this paper, we establish mathematical relationships between the intrinsic growth rate \( r \) in \eqref{eq:logistic3} and \eqref{ricker1} and the parameters of the gamma distribution, while analyzing the influence of the variance of \( \epsilon_t \) on equilibrium states. Notably, the mathematical formulation reveals the existence of two branches of the intrinsic growth rate, \( r_+ \) and \( r_- \), which represent alternative stable states corresponding to higher and lower growth rates, respectively. Furthermore, the intrinsic growth rate \( r \) is shown to be independent of \( \theta \), the scale parameter of the gamma distribution. This indicates that \( r \) is primarily governed by the population's internal dynamics, as reflected by \( k \), rather than by the scale or spread of the population size distribution dictated by \( \theta \). This highlights the critical role of internal biological mechanisms in determining growth, irrespective of the absolute population size or its variability.

This work extends the analysis of the stochastic logistic equation presented in \cite{Wang2024} by addressing more complex logistic equations with power nonlinearities and the Ricker equation. The incorporation of these advanced models provides deeper insights into the interplay between stochastic effects, population growth rates, and distributional parameters, offering a broader framework for understanding real-world population dynamics.

Our key contributions in this paper are as follows:

\begin{enumerate}
    \item \textbf{Mathematical Relationships}: We establish the mathematical relationship between the intrinsic growth rate \( r \) in \eqref{eq:logistic3} and \eqref{ricker1} and the parameters of the gamma distribution. This includes investigation of the influence of the perturbation \( \epsilon_t \) on \( r \), providing theoretical insights into how the stochastic models compare to their deterministic counterparts in describing population dynamics.

    \item \textbf{Alternative Stable States}: We identify two branches of the intrinsic growth rate, \( r_+ \) and \( r_- \), for \eqref{eq:logistic3} and \eqref{ricker1}. These branches represent alternative stable states, corresponding to higher and lower growth rates, respectively. This duality provides deeper insights into population stability and resilience under stochastic conditions.

    \item \textbf{Biological Implications}: We analyze the biological significance of these relationships, with a focus on how the shape parameter \( k \) and scale parameter \( \theta \) of the gamma distribution influence population dynamics at equilibrium. This analysis highlights the role of internal biological mechanisms and environmental variability in shaping the behavior of populations under stochastic effects.
\end{enumerate}

\section{Equilibrium Analysis of Modified Discrete Stochastic Logistic Equation}
\label{sec:modified_equation}

\subsection{The Gamma Distribution at Equilibrium}
The gamma distribution, on the interval \( (0, \infty) \), has two key parameters: the shape parameter \( k \) and the scale parameter \( \theta \). Its probability density function (PDF) is expressed as:

\begin{equation} \label{eq:gamma_pdf}
f(x; k, \theta) = \frac{x^{k - 1} e^{-x / \theta}}{\Gamma(k) \theta^k}, \quad \text{for } x > 0,
\end{equation}
where \( \Gamma(k) \) is the gamma function, a generalization of the factorial for real and complex arguments, with \( \Gamma(n) = (n-1)! \) for positive integers \( n \). The parameter \( k \), often called the shape parameter, determines the form of the distribution, including its skewness and tail behavior. The scale parameter \( \theta \) influences the spread, with larger values of \( \theta \) producing broader distributions  \cite{wikipedia_gamma}.

The gamma distribution is highly versatile and has been applied in a variety of fields. In population dynamics, it models population size distributions, capturing the variability and skewness inherent in biological systems. In stochastic processes and event modeling, it is commonly used to describe waiting times between events, as it naturally fits data that is positively skewed. Its mathematical properties and flexibility make it a critical tool for analyzing ecological systems, biological processes, and other phenomena characterized by asymmetry and variability \cite{wikipedia_gamma}.

The gamma distribution has theoretical significance as it often emerges as a stationary solution to stochastic processes which incorporate random fluctuations into population dynamics, such as birth-death models and stochastic differential equations (SDEs).  These stochastic models account for variability around a deterministic growth process, providing a more realistic framework for representing natural populations. At equilibrium, the statistical properties of the population stabilize over time, with the gamma distribution effectively capturing this behavior. Its capacity to describe populations that settle around an equilibrium point under the influence of stochastic effects makes it a valuable tool for modeling real-world population dynamics. Numerous studies have demonstrated that the gamma distribution serves as an accurate approximation of stationary distributions in diverse ecological contexts \cite{Dennis1984, Peters1989, DennisCostantino1988, Costantino1981}.

We assume that the population size $X_t$ in \eqref{eq:logistic3} at equilibrium follows a gamma distribution with parameters $k$ and $\theta$:
\begin{equation} \label{eq:x_gamma_distribution}
X_t \sim \text{Gamma}(k, \theta).
\end{equation}
Because of the nonlinearity and random perturbation in (\ref{eq:logistic3}), $X_{t+1}$ may not necessarily follow the gamma distribution. However, in order to study relations of these parameters in the models, it is reasonable to assume that the population at equilibrium from specific time $t$ to $t+1$ maintains the same expectation and variance. Therefore, at equilibrium, we assume that, for a specific time $t$,  
\begin{equation} \label{eq:conditions}
E[X_{t+1}] = E[X_t], \,\,\,  Var[X_{t+1}] = Var[X_t] 
\end{equation}
which allows us to derive mathematical relations for $r$ in terms of $k$ and $\theta$. 

\subsection{Mean Condition at Equilibrium}
\label{subsec:mean_equilibrium}
At equilibrium, we have \( E[X_{t+1}] = E[X_t] = \mu \). It is known from the identity \eqref{identity} that the first two moments are:
\begin{align}
E[X_t] &= \mu = k \theta, \label{eq:mean_mu} \\
E[X_t^2] &= k (k + 1) \theta^2. \label{eq:second_moment}
\end{align}
and the mean and variance of the gamma distribution are:
\begin{align}
\mu &= E[X_t] = k \theta, \label{eq:gamma_mean} \\
\sigma^2 &= \text{Var}[X_t] = k \theta^2. \label{eq:gamma_variance}
\end{align}
The higher order moments of \( X_t \) are:
\begin{align}
E[X_t^n] &= \theta^n \frac{\Gamma(k+n )}{\Gamma(k)}, \label{eq:gamma_moment} \\
\end{align}
Since $E[\epsilon_t]=1$, we computer \( E[X_{t+1}] \):

\begin{align}
E[X_{t+1}] &= E\left[ X_t \left( r - r X_t^n \right) \epsilon_t \right] \nonumber \\
&= r E[X_t] - r E[X_t^{n+1}]. \label{eq:E_Xt1}
\end{align}
Setting \( E[X_{t+1}] = E[X_t] = \mu \) gives
\begin{equation}
\mu (1 - r) + r E[X_t^{n+1}] = 0. \label{eq:mean_eq_rearranged}
\end{equation}
Substituting \( \mu = k \theta \) and \( E[X_t^{n+1}] = \theta^{n+1} \frac{\Gamma(k + n + 1)}{\Gamma(k)} \) gives

\begin{equation}
k \theta (1 - r) + r\theta^{n+1} \frac{\Gamma(k + n + 1)}{\Gamma(k)} = 0. \label{eq:mean_eq_substituted}
\end{equation}
Divide both sides by \( k \theta \):
\begin{equation}
(1 - r) + \frac{r\theta^{n}}{ k} \frac{\Gamma(k + n + 1)}{\Gamma(k)} = 0. \label{eq:mean_eq_divided}
\end{equation}
It follows that 
\begin{equation}
\theta^{n} = \frac{k (r - 1)}{r} \frac{\Gamma(k)}{\Gamma(k + n + 1)}. \label{eq:y_over_K}
\end{equation}
It follows that $r>1$ must hold at equilibrium of \eqref{eq:logistic3}.
\subsection{Variance Condition at Equilibrium}
\label{subsec:variance_equilibrium}
The assumption  \( \text{Var}(X_{t+1}) = \text{Var}(X_t) \) will provide additional conditions that can help determine the gamma distribution parameters.  To compute  \( \text{Var}(X_{t+1}) \), we need \( E[X_{t+1}^2] \), given \( E[X_{t+1}]=E[X_{t}] \) and 
\[
\text{Var}(X_{t+1}) = E[X_{t+1}^2] - \left( E[X_{t+1}] \right)^2.
\]
Since \( \text{Var}(X_{t+1}) = \text{Var}(X_t) \), we have 
\begin{equation}
E[X_{t}^2] = E[r^2X_t^2(1-X_t^n)^2 \epsilon_t^2] =\left (r^2 E[X_t^2] - 2 r^2 E[X_t^{n+2}] + r^2 E[X_t^{2n+2}] \right) E[\epsilon_t^2]. \label{eq:E_Xt1_squared}
\end{equation}
Divide by $-E[\epsilon_t^2]$ and add \( E[X_t^2] \) to both sides:
\begin{equation}
E[X_t^2] - r^2 E[X_t^2] + 2 r^2 E[X_t^{n+2}] - r^2 E[X_t^{2n+2}] = \frac{E[\epsilon_t^2]-1}{E[\epsilon_t^2]} E[X_t^2].
\end{equation}
and
\begin{equation}
(1 - r^2) E[X_t^2] + 2 r^2 E[X_t^{n+2}] - r^2 E[X_t^{2n+2}] = \frac{\text{Var}(\epsilon_t)}{\text{Var}(\epsilon_t)+1} E[X_t^2]. \label{eq:variance_eq_simplified2}
\end{equation}
Now, express the moments using the gamma distribution:

\begin{align}
E[X_t^2] &= \theta^2 \frac{\Gamma(k + 2)}{\Gamma(k)} = \theta^2 k (k + 1), \label{eq:E_Xt2} \\
E[X_t^{n+2}] &= \theta^{n+2} \frac{\Gamma(k + n + 2)}{\Gamma(k)}, \label{eq:E_Xt_n+2} \\
E[X_t^{2n+2}] &= \theta^{2n+2} \frac{\Gamma(k + 2n + 2)}{\Gamma(k)}. \label{eq:E_Xt_2n+2}
\end{align}
Substitute these expressions into the variance equation \eqref{eq:variance_eq_simplified2}:

\begin{equation}
(1 - r^2) \theta^2 k (k + 1) + 2 r^2 \theta^{n+2} \frac{\Gamma(k + n + 2)}{\Gamma(k)} - r^2 \theta^{2n+2} \frac{\Gamma(k + 2n + 2)}{\Gamma(k)} = \frac{\text{Var}(\epsilon_t)}{\text{Var}(\epsilon_t)+1} \theta^2 k (k + 1). \label{eq:variance_eq_with_moments}
\end{equation}
Divide both sides by \( \theta^2 \):

\begin{equation}
(1 - r^2) k (k + 1) + 2 r^2 \theta^{n} \frac{\Gamma(k + n + 2)}{\Gamma(k)} - r^2 \theta^{2n} \frac{\Gamma(k + 2n + 2)}{\Gamma(k)} = \frac{\text{Var}(\epsilon_t)}{\text{Var}(\epsilon_t)+1} k (k + 1). \label{eq:variance_eq_divided1}
\end{equation}
Let 
\[
G = \frac{\Gamma(k)}{\Gamma(k + n + 1)}
\]
From equation \eqref{eq:y_over_K},
\[
\theta^{n} = k \left( \frac{r - 1}{r} \right) G
\]
Thus, substituting \( \theta^{n} \) into equation \eqref{eq:variance_eq_divided1}, we get:
\begin{align}
& (1 - r^2) \, k (k + 1) + 2 r^2 \, \left[ k \left( \frac{r - 1}{r} \right) G \right] \, \frac{\Gamma(k + n + 2)}{\Gamma(k)} \nonumber \\
& \quad {} - r^2 \, \left[ k \left( \frac{r - 1}{r} \right) G \right]^2 \, \frac{\Gamma(k + 2n + 2)}{\Gamma(k)} = \frac{\text{Var}(\epsilon_t)}{\text{Var}(\epsilon_t)+1} \, k (k + 1).
\label{eq:substituted}
\end{align}
which can be simplified:
\begin{align}
(1 - r^2) \, k (k + 1) + 2 r (r - 1) k G \, \frac{\Gamma(k + n + 2)}{\Gamma(k)} - (r - 1)^2 k^2 G^2 \, \frac{\Gamma(k + 2n + 2)}{\Gamma(k)} \nonumber \\
= \frac{\text{Var}(\epsilon_t)}{\text{Var}(\epsilon_t)+1} \, k (k + 1).
\label{eq:expanded}
\end{align}
Next, we divide both sides by \( k (r - 1) \) and we have:
\begin{align}
- (1 + r)(k + 1) + 2 r G \, \frac{\Gamma(k + n + 2)}{\Gamma(k)} - (r - 1) k G^2 \, \frac{\Gamma(k + 2n + 2)}{\Gamma(k)} \nonumber \\
= \frac{\text{Var}(\epsilon_t)}{\text{Var}(\epsilon_t)+1} \, \frac{k + 1}{r - 1}.
\label{eq:simplified3}
\end{align}
Let
\[
Q =  G \cdot \frac{\Gamma(k + 2n + 2)}{\Gamma(k + n + 1)}.
\]
which is
\begin{equation}
Q =  \frac{\Gamma(k)}{\Gamma(k + n + 1)} \frac{\Gamma(k + 2n + 2)}{\Gamma(k + n + 1)}  = \frac{\Gamma(k + 2n + 2) \cdot \Gamma(k)}{\left[ \Gamma(k + n + 1) \right]^2}
\label{eq:simplified_q}
\end{equation}
Now, the equation \eqref{eq:simplified3} becomes:
\begin{align}
- (1 + r)(k + 1) + 2 r (k + n + 1) - (r - 1) k Q = \frac{\text{Var}(\epsilon_t)}{\text{Var}(\epsilon_t)+1} \, \frac{k + 1}{r - 1}.
\label{eq:with_Q1}
\end{align}
Rearranging the left side of \eqref{eq:with_Q1} gives
\[
(r - 1)(k + 1) - (r - 1) k Q + 2 r n = \frac{\text{Var}(\epsilon_t)}{\text{Var}(\epsilon_t)+1} \, \frac{k + 1}{r - 1}.
\]
and 
\begin{equation}
(r - 1) \left[ k (1 - Q) + 1 \right] + 2 r n = \frac{\text{Var}(\epsilon_t)}{\text{Var}(\epsilon_t)+1} \, \frac{k + 1}{r - 1}.
\label{eq:factored}
\end{equation}
Multiplying both sides by \( r - 1 \), we have:
\begin{equation}
(r - 1)^2 \left[ k (1 - Q) + 1 \right] + 2 r n (r - 1) = \frac{\text{Var}(\epsilon_t)}{\text{Var}(\epsilon_t)+1} \, (k + 1).
\label{eq:factored1}
\end{equation}
Let:
\[
A = k(1 - Q) + 1,
\]
\[
B = \frac{\text{Var}(\epsilon_t)}{\text{Var}(\epsilon_t)+1} \, (k + 1).
\]
and we have
\begin{equation}
(r - 1)^2 A + 2 r n (r - 1) = B
\label{eq:simplified23}
\end{equation}
which is
\begin{equation}
(A + 2 n) r^2 - (2 A + 2 n) r + (A - B) = 0.
\label{eq:simplified233}
\end{equation}
The discriminant \( D \) of \eqref{eq:simplified233} is:
\begin{align*}
D &= \left[ -2 (A + n) \right]^2 - 4 (A + 2 n)(A - B), \\
&= 4 (A + n)^2 - 4 (A + 2 n)(A - B), \\
&= 4 \left[ (A + n)^2 - (A + 2 n)(A - B) \right]\\
&= 4 \left( n^2 + A B + 2 n B \right).
\end{align*}
Using the quadratic formula, we find the solution of \eqref{eq:simplified233}
\begin{align*}
r &= \frac{ - \left( -2 (A + n) \right) \pm \sqrt{D} }{ 2 (A + 2 n) }, \\
&= \frac{ 2 (A + n) \pm \sqrt{D} }{ 2 (A + 2 n) }, \\
&= \frac{ A + n \pm \dfrac{\sqrt{D}}{2} }{ A + 2 n }.
\end{align*}
Thus, the general solution for \( r \) in  \eqref{eq:simplified233} or \eqref{eq:factored1} is:
\begin{equation}
r_{\pm} = \frac{ A + n \pm \sqrt{ n^2 + A B + 2 n B } }{ A + 2 n }.
\label{eq:solution_r2}
\end{equation}

Let's verify it when \( n = 1 \). When $n=1$
\[
Q = \frac{\Gamma(k + 4) \cdot \Gamma(k)}{\left[ \Gamma(k + 2) \right]^2}=\frac{ (k + 2)(k + 3) }{ k (k + 1) }
\]
$$
A= k(1 - Q) + 1=\dfrac{ -3 k - 5 }{ k + 1 }
$$
and therefore 
\begin{equation}
r = \frac{ A + 1 \pm \sqrt{ 1 + (A +2)B } }{ A + 2  }=\frac{ -\frac{2k+4}{k+1} \pm \sqrt{ 1 -\frac{k+3}{k+1}B} }{ -\frac{k+3}{k+1}}==\frac{ 2k+4 \pm (k+1)\sqrt{ 1 -\frac{k+3}{k+1}B} }{ k+3}.
\label{eq:solution_r}
\end{equation}
and 
\begin{equation}
r = \frac{ 2k+4 \pm (k+1)\sqrt{ \dfrac{1 - Var[\epsilon_t](k + 2)}{Var[\epsilon_t] + 1} } }{k+3}.
\label{eq:solution_r1}
\end{equation}
This is exactly the formula in \cite{Wang2024} for the stochastic logistic equation when $n=1$.

\subsection{Biological Interpretation}
For $n=1$, the explicit expression \eqref{eq:solution_r1} holds. \cite{Wang2024} shows that the parameters must satisfy $$0 \leq Var[\epsilon_t] \leq 0.5,$$ and for each $Var[\epsilon_t]$ within the range,  $r$ exists only for  $$ 0< k \leq \dfrac{1}{Var[\epsilon_t]}-2.$$

\begin{figure}[h!]
    \centering
    \begin{subfigure}[b]{0.45\textwidth}
        \centering
        \includegraphics[width=\textwidth]{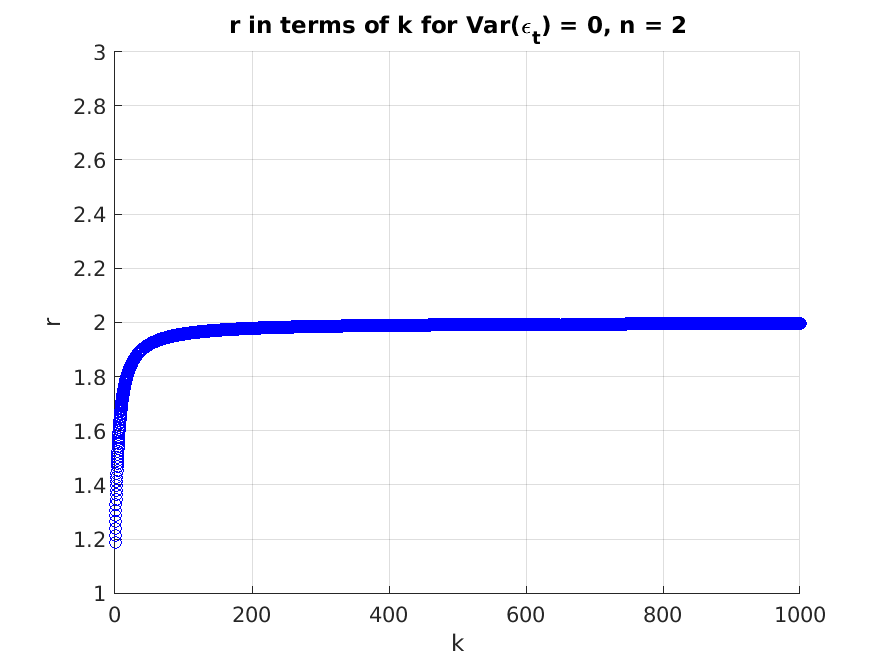}
%        \caption{Caption 1}
        \label{fig:sub1}
    \end{subfigure}
    \hfill
    \begin{subfigure}[b]{0.45\textwidth}
        \centering
        \includegraphics[width=\textwidth]{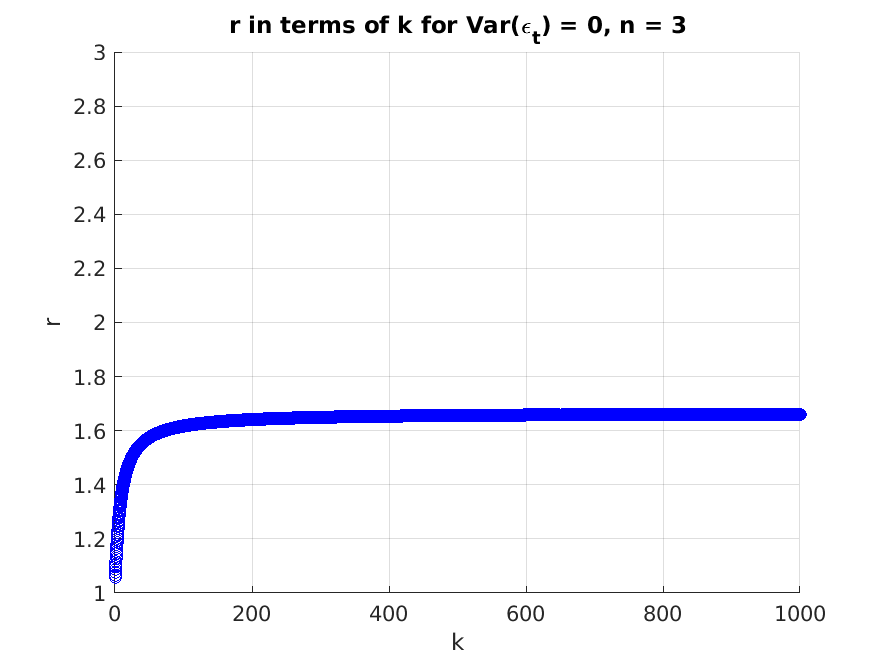}
%        \caption{Caption 2}
        \label{fig:sub2}
    \end{subfigure}
    
    \vskip\baselineskip
    
    \begin{subfigure}[b]{0.45\textwidth}
        \centering
        \includegraphics[width=\textwidth]{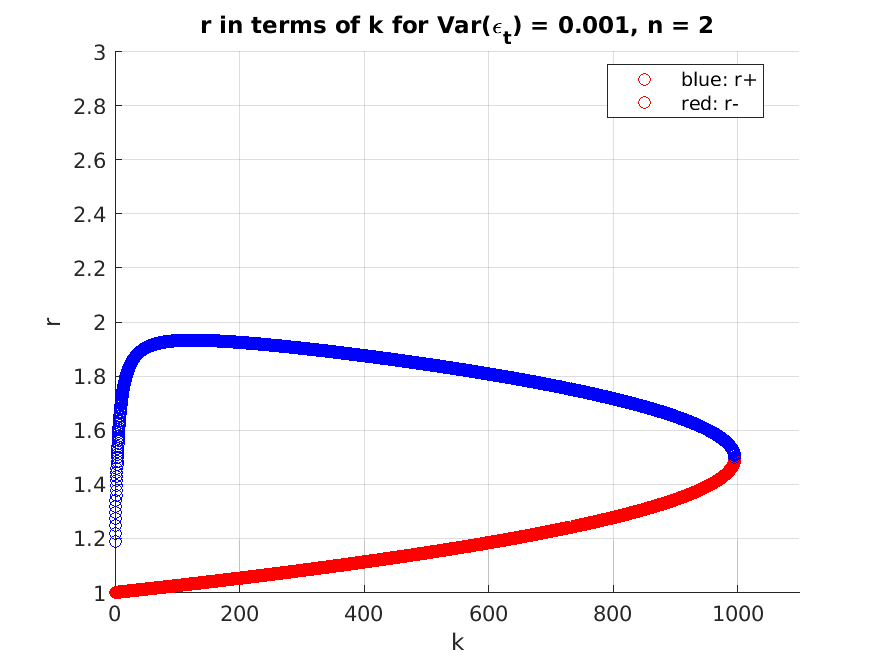}
 %       \caption{Caption 3}
        \label{fig:sub3}
    \end{subfigure}
    \hfill
    \begin{subfigure}[b]{0.45\textwidth}
        \centering
        \includegraphics[width=\textwidth]{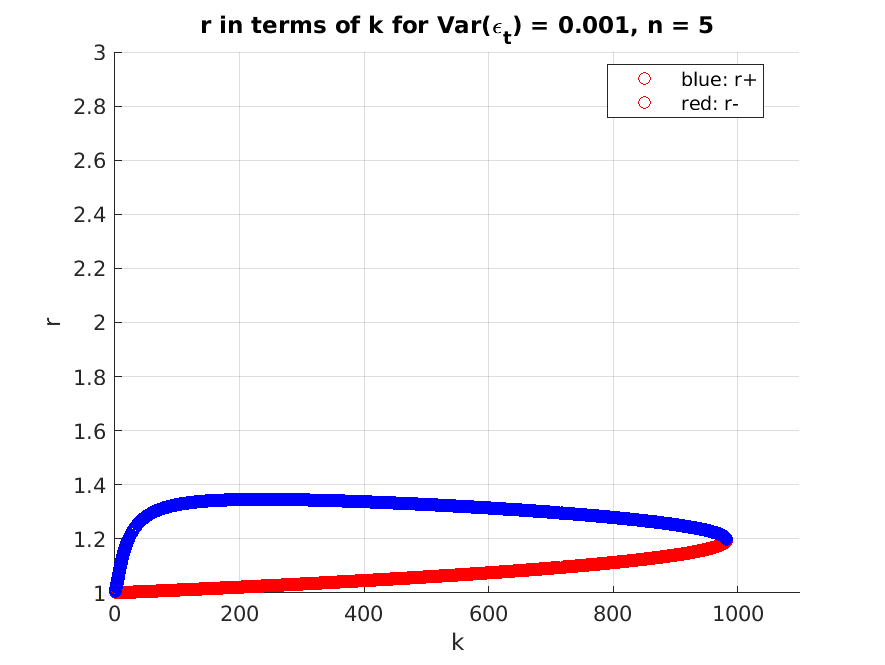}
  %      \caption{Caption 4}
        \label{fig:sub4}
    \end{subfigure}
    
    \vskip\baselineskip
    
    \begin{subfigure}[b]{0.45\textwidth}
        \centering
        \includegraphics[width=\textwidth]{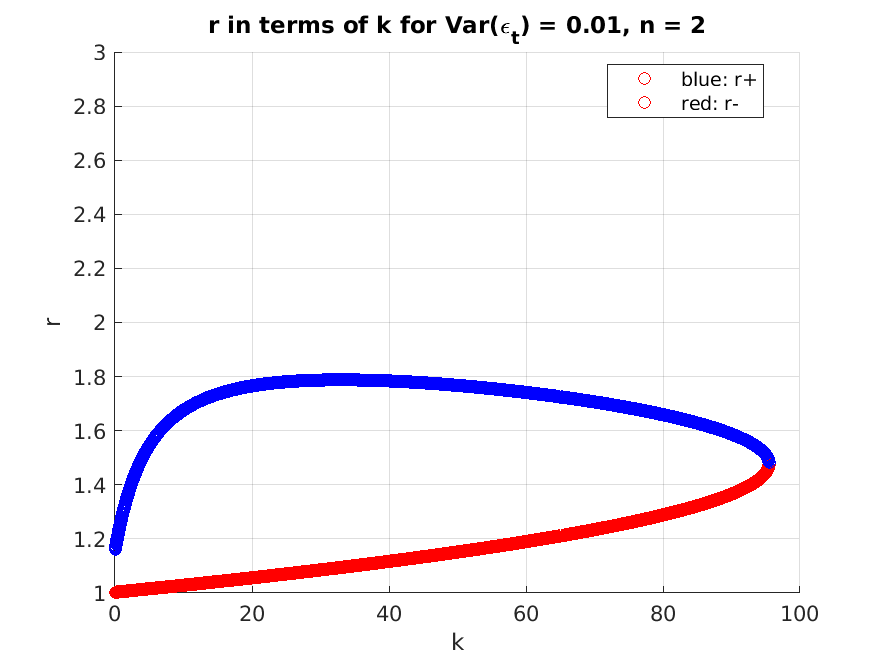}
 %       \caption{Caption 3}
        \label{fig:sub3}
    \end{subfigure}
    \hfill
    \begin{subfigure}[b]{0.45\textwidth}
        \centering
        \includegraphics[width=\textwidth]{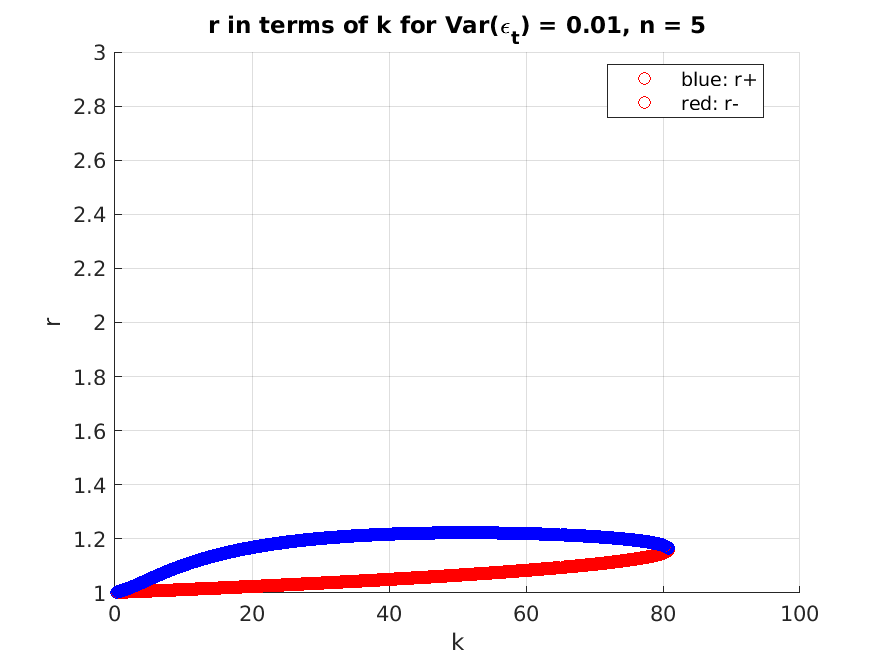}
        \caption{Caption 4}
        \label{fig:sub4}
    \end{subfigure}   
    \caption{Plots of $r$ in term of $k$ for the modified logistic equation}
    \label{fig:main}
\end{figure}

For $n>1$, it is challenging to give explicit expressions for upper bounds of $\text{Var}(\epsilon_t)$ and $k$ as for $n=1$. Nevertheless, it is easy to see that, in the expression $r$ in terms of $k$ in \eqref{eq:solution_r2}, $B = \frac{\text{Var}(\epsilon_t)}{\text{Var}(\epsilon_t)+1} \, (k + 1)>0$ and only $A$ may be negative, and      $\frac{\text{Var}(\epsilon_t)}{\text{Var}(\epsilon_t)+1}$ is increasing with respect to $\text{Var}(\epsilon_t)$.  Now in order that the discriminant $D \geq 0$, there is an upper bound of $\text{Var}(\epsilon_t)$ depending on $n$, and for each $n$ and corresponding $\text{Var}(\epsilon_t)>0$, there is a upper bound for $k$ such that $r$ has two solutions as depicted in Figure \ref{fig:main}. We would like to see its impact on $r$ through simulations in Figure \ref{fig:main}. The four figures illustrate the relationship between the growth rate \( r \) and the parameter \( k \) under different values \( \text{Var}(\epsilon_t) \). 
  
First, we note that all values of \( r \) in Figure \ref{fig:main} are within the range \( (1, 3) \), as expected for the both deterministic and stochastic logistic models. In particular, for the deterministic logistic model, populations grow and stabilize in the same range, where the growth rate balances with the carrying capacity, allowing the population to approach a stable state \cite{may1976simple,strogatz2018nonlinear,hilborn2000chaos}.

Figure \ref{fig:main} illustrates that there are two solutions for \( r \) in \eqref{eq:solution_r2}, namely \( r_+ \) and \( r_- \), representing alternative stable states of the intrinsic growth rate \( r \) under \( \text{Var}(\epsilon_t) > 0 \). For \( \text{Var}(\epsilon_t) = 0 \), the trivial solution \( r = 1 \) always exists. The branch \( r_+ \) typically corresponds to higher growth rates, indicating a more resilient population and suggesting a higher-density equilibrium. This state is likely stable for species thriving in resource-abundant environments. Conversely, the branch \( r_- \) represents lower growth rates, corresponding to populations that may be more vulnerable to fluctuations. Populations at \( r_- \) are indicative of a low-density equilibrium, potentially more sensitive to environmental variability and at greater risk of extinction in the presence of high environmental variance.

For \( \text{Var}[\epsilon_t] > 0 \), the feasible range of \( k \) values is constrained by the environmental variance, as illustrated in Figure \ref{fig:main}. First, as \( n \) increases, the growth rate \( r \) decreases due to the higher-order death term \( X^n \), which affects a larger portion of the population, reducing its overall growth potential. Smaller values of \( \text{Var}(\epsilon_t) \) expand the range of \( k \), suggesting that populations with a larger gamma shape parameter \( k \) are better supported in stable environments with low variability. In general, from the definition of the gamma distribution, it is easy to see that a population with a larger gamma shape parameter has a more symmetrical and regular population distribution. From Figure \ref{fig:main}, we observe that \( r \) generally increases with respect to \( k \), which is particularly true for the low branch, $r_{-}$. Consequently, lower \( \text{Var}(\epsilon_t) \) implies greater potential for a stable equilibrium across a wider range of \( r \). The shape parameter \( k \) thus plays a critical role in defining equilibrium conditions for populations in varying environments.

In Figure \ref{fig:main}, for smaller values of \( \text{Var}(\epsilon_t) \), such as \( \text{Var}(\epsilon_t) = 0.0001, n=2 \) and \( \text{Var}(\epsilon_t) = 0.0001, n=5 \), the feasible range of \( k \) is extensive and up to 1000. In this context, populations are expected to exhibit a diverse range of equilibrium growth rates, represented by the \( r_+ \) and \( r_- \) branches. Higher \( k \) values under low variance indicate a less skewed and more symmetrical population distribution, reflecting greater resilience and stability. This setup supports the potential for a stable equilibrium across a wider range of \( r \) in stable environments with minimal environmental variability. Such conditions enable these populations to sustain predictable dynamics and maintain consistency around their equilibrium states.

In Figure \ref{fig:main}, as \( \text{Var}(\epsilon_t) \) increases (e.g., \( \text{Var}(\epsilon_t) = 0.01 \) for \( n = 2 \) and \( n = 5 \)), the range of \( k \) values supporting stable solutions becomes narrower. This suggests that populations in highly variable environments are more constrained in terms of the shape parameter \( k \), limiting their ability to maintain stable equilibria.

The independence of \( r \) from \( \theta \) highlights that the intrinsic growth rate is determined primarily by the population's internal dynamics, represented by \( k \), rather than by the scale of the population size distribution, which is governed by \( \theta \) in the gamma distribution. This underscores the significance of internal biological processes in driving population growth, irrespective of the absolute size of the population or the variability in its distribution.

\section{Equilibrium analysis of Stochastic Ricker Equation}

\subsection{Equilibrium Analysis with Gamma Distribution}
In this section, we aim to derive the mean and variance equilibrium conditions of \eqref{ricker1} under the assumption that \( X_t \) follows a gamma distribution at equilibrium. At equilibrium, \( X_t \) follows a gamma distribution with shape parameter \( k \) and scale parameter \( \theta \):

\[
X_t \sim \text{Gamma}(k, \theta)
\]
Although $X_{t+1}$ may not follow a gamma distribution, it is reasonable to assume that the mean and variance are the same: 
\[
E[X_{t+1}] = E[X_t] = \mu \quad \text{and} \quad \text{Var}(X_{t+1}) = \text{Var}(X_t)
\]
Note that, because $E[\epsilon_t]=1$ 
\begin{align}
E[X_{t+1}] &= E\left[ X_t e^{r(1 - X_t)} \right]E[\epsilon_t] \\
&= e^{r} E\left[ X_t e^{- r X_t} \right] 
\end{align}
In order to compute \( E\left[ X_t e^{- r X_t} \right] \), for a gamma-distributed variable \( X_t \sim \text{Gamma}(k, \theta) \), we can use the following identity \eqref{identity}. Its proof is provided in the Appendix 

\begin{equation}
E\left[ X_t^n e^{- s X_t} \right] = \frac{\Gamma(k + n)}{\Gamma(k)} \frac{\theta^n}{(1 + s \theta)^{k + n}}
\label{mean_identity}
\end{equation}
Setting \( s = r, n=1 \), we have:
\[
E\left[ X_t e^{- r X_t} \right] = \frac{k \theta}{(1 + r \theta)^{k + 1}}
\]
Therefore, the mean equilibrium condition becomes:
\begin{align}
k \theta &= e^{r} \frac{k \theta}{(1 + r \theta)^{k + 1}}
\end{align}
Divide both sides by \( k \theta \) and rearrange the terms:
\begin{equation}
e^{- r} = \left( \frac{1}{1 + r \theta} \right)^{k + 1}
\end{equation}
Inverting both sides gives
\begin{equation}
\left( 1 + r \theta \right)^{k + 1} = e^{r}
\label{mean_equilibrium}
\end{equation}

We now turn to the variance condition. Let's compute \( E[X_{t+1}^2] \):

\begin{align}
E[X_{t+1}^2] &= E\left[ \left( X_t e^{r(1 - X_t)} \epsilon_t \right)^2 \right] \\
&= e^{2 r} E\left[ X_t^2 e^{- 2 r X_t} \right] E[\epsilon_t^2]
\end{align}
Using \eqref{mean_identity}, for \( n = 2 \) and \( s = 2 r \), we have 

\begin{equation}
E[X_{t+1}^2] = e^{2 r} k (k + 1) \theta^2 \left( \frac{1}{1 + 2 r \theta} \right)^{k + 2} E[\epsilon_t^2]
\end{equation}
At equilibrium, we assume that 

\[
\text{Var}(X_{t+1}) = \text{Var}(X_t) = k \theta^2
\]
Because 
\[
\text{Var}(X_{t+1}) = E[X_{t+1}^2] - (E[X_{t+1}])^2
\]
we have 
\[
E[X_{t+1}^2] = k \theta^2 + (k \theta)^2 = k \theta^2 (1 + k)
\]
Now we arrive at 
\begin{equation}
e^{2 r} k (k + 1) \theta^2 \left( \frac{1}{1 + 2 r \theta} \right)^{k + 2}E[\epsilon_t^2] = k \theta^2 (1 + k)
\end{equation}
Dividing both sides by \( k \theta^2 (k + 1) \) gives 
\begin{equation}
\left( 1 + 2 r \theta \right)^{k + 2} = e^{2 r} E[\epsilon_t^2]
\label{variance_equilibrium1}
\end{equation}
From Equation \eqref{mean_equilibrium}:
\begin{equation}
1 + r \theta = e^{\frac{r}{k + 1}}
\label{theta_from_mean}
\end{equation}
From Equation \eqref{variance_equilibrium1}:
\begin{equation}
1 + 2 r \theta = (E[\epsilon_t^2] e^{2r})^{\frac{1}{k + 2}}
\label{theta_from_variance}
\end{equation}
Eliminating \( \theta \) from both equations gives
\begin{equation}
\frac{1}{r} \left( e^{\frac{r}{k + 1}} - 1 \right) = \frac{1}{2 r} \left( (E[\epsilon_t^2] e^{2r})^{\frac{1}{k + 2}}- 1 \right)
\end{equation}
and 
\begin{equation}
2 e^{\frac{r}{k + 1}} - (E[\epsilon_t^2] e^{2r})^{\frac{1}{k + 2}} = 1
\label{final_equation}
\end{equation}
Now we have derived a transcendental equation involving \( r \) and \( k \):

\begin{equation}
2 e^{\frac{r}{k + 1}} - (E[\epsilon_t^2] e^{2r})^{\frac{1}{k + 2}} = 1
\end{equation}
That is 
\begin{equation}
2 e^{\frac{r}{k + 1}} - \left ((1+Var(\epsilon_t)) e^{2r}\right )^{\frac{1}{k + 2}} = 1
\label{final_equation1}
\end{equation}
This equation can be solved numerically for specific values of \( k \) to find the relationship between \( r \) and \( k \) at equilibrium.

\subsection{Biological Interpretation}
\begin{figure}[h!]
    \centering
    \begin{subfigure}[b]{0.45\textwidth}
        \centering
        \includegraphics[width=\textwidth]{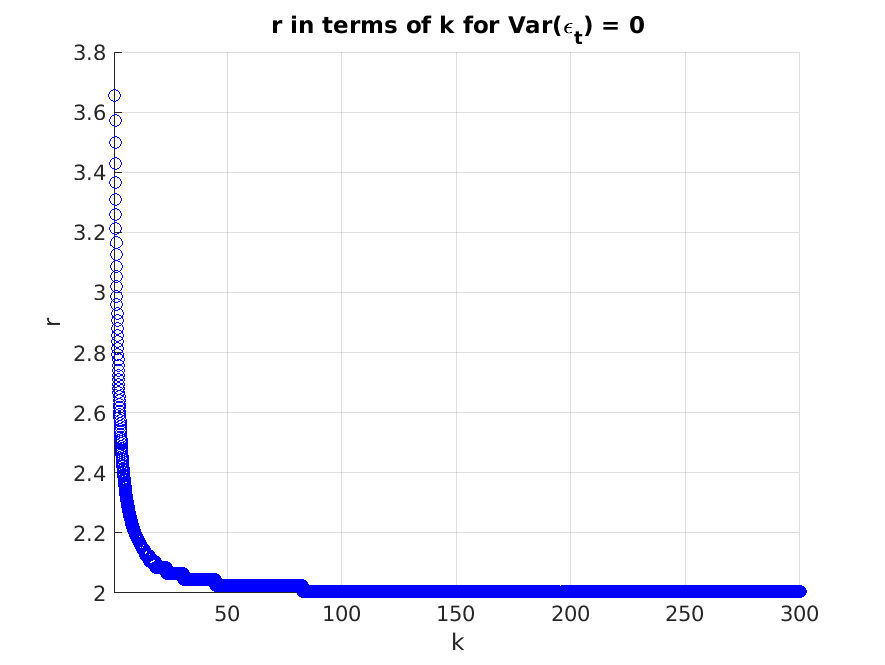}
        \label{fig:sub1}
    \end{subfigure}
    \hfill
    \begin{subfigure}[b]{0.45\textwidth}
        \centering
        \includegraphics[width=\textwidth]{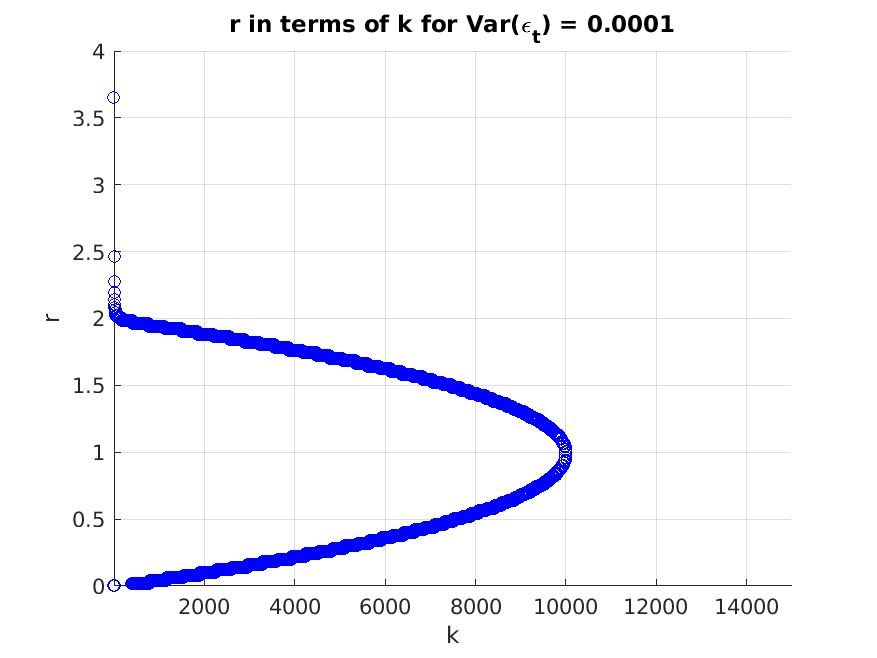}
%        \caption{Caption 2}
        \label{fig:sub2}
    \end{subfigure}
    
    \vskip\baselineskip
    
    \begin{subfigure}[b]{0.45\textwidth}
        \centering
        \includegraphics[width=\textwidth]{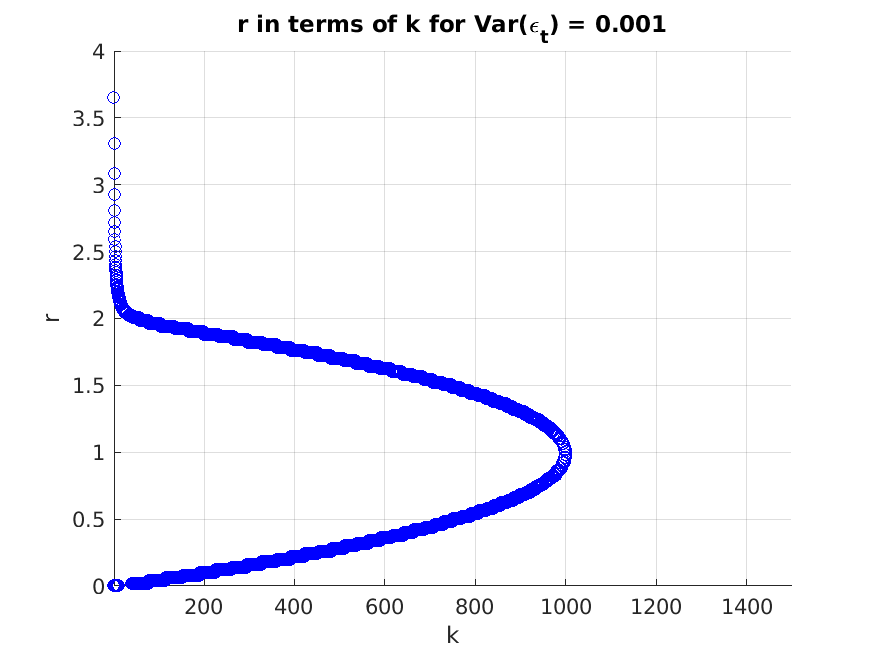}
 %       \caption{Caption 3}
        \label{fig:sub3}
    \end{subfigure}
    \hfill
    \begin{subfigure}[b]{0.45\textwidth}
        \centering
        \includegraphics[width=\textwidth]{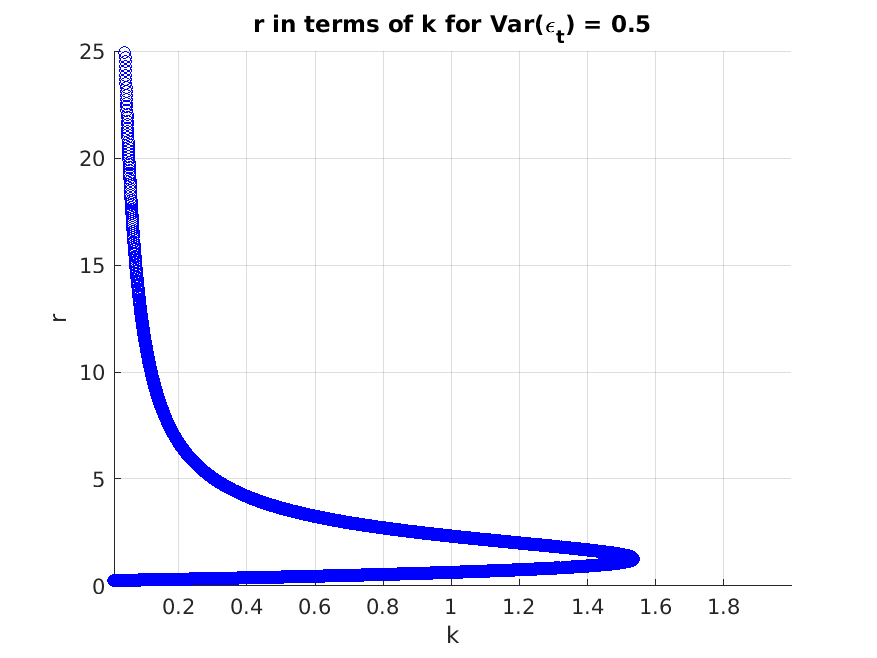}
  %      \caption{Caption 4}
        \label{fig:sub4}
    \end{subfigure}
        
    \caption{Plots of $r$ in term of $k$ for Ricker equation}
    \label{fig:main1}
\end{figure}

In this section, we use numerical simulations to study the relation between $r$ and $k$ in \ref{final_equation1} for stochastic Ricker equation \eqref{ricker1}. We would like to see the impact of $Var(\epsilon_t)$ on $r$ through simulations in Figure \ref{fig:main1}. The four figures illustrate the relationship between the growth rate \( r \) and the parameter \( k \) under different values \( \text{Var}(\epsilon_t) \). 
    
For the deterministic Ricker model given by \eqref{ricker1},  population dynamics, including stable growth, periodic oscillations, chaos, or extinction, are closely tied to the value of \( r \) \cite{may1976simple,kot2001elements}. For \( 0 < r < 2.69 \), the population typically grows and stabilizes at a nonzero steady state or exhibits damped oscillations or populations alternating between large and small sizes in successive generations. For \( r > 2.69 \), the system can enter a chaotic regime, characterized by unpredictable and irregular fluctuations \cite{may1976simple,kot2001elements}. In contrast, in the stochastic version of \eqref{ricker1}, as we see from Figure \ref{fig:main1} while the lower branch  of \( r \) is less than $2$ at equilibrium, but the upper branch may exceed 2.69 for $k$ is small, reflecting the stabilizing influence of stochastic effects.  Smaller values of $k$ correspond to highly skewed distributions.  populations with low $k$ values are more susceptible to environmental fluctuations. 
  
For $\text{Var}(\epsilon_t)>0$,  there are two branches of $r$. The upper bounds of $k$ are determined by $\text{Var}(\epsilon_t)$. The upper branch consistently corresponds to higher growth rates and represents populations in resource-abundant or favorable conditions. These populations are more likely to achieve higher densities and maintain stability even under mild environmental perturbations. The lower branch reflects reduced growth rates, often associated with low-density populations that are more susceptible to environmental fluctuations with less risk for chaotic behaviors.

When the variance \( \text{Var}(\epsilon_t)=0 \), as shown in the first figure in Figure \ref{fig:main1}), $r$ has only one positive branch ($r=0$ is always a trivial branch).  The transition from a single positive branch (\( \text{Var}(\epsilon_t) = 0 \)) to two branches (\( \text{Var}(\epsilon_t) > 0 \)) marks a significant shift in system dynamics. This bifurcation indicates the introduction of stochastic perturbation effects that alter the equilibrium conditions and diversify the potential population outcomes. As \( \text{Var}(\epsilon_t) \) increases, the bifurcation points shift, leading to narrower ranges of \( k \) and particular for the lower branch, $r$ approaches values leading to more stable dynamics.

In environments with low variance (\( \text{Var}(\epsilon_t) \ll 1 \)), populations can sustain higher \( k \) values and maintain relatively stable dynamics across a wide range of \( r \). These conditions are conducive to long-term persistence and robust population equilibria.
For high variance (\( \text{Var}(\epsilon_t) \approx 0.5 \)), populations face significant instability. The restricted range of \( k \) and the convergence of the two branches suggest that populations are more likely to experience irregular fluctuations, reduced densities, or even extinction. 

The figures in Figure \ref{fig:main1} provide a comprehensive view of how growth rates (\( r \)), population distributions (\( k \)), and environmental variance (\( \text{Var}(\epsilon_t) \)) interact to shape population dynamics. These insights are crucial for predicting and managing population stability in fluctuating environments.

\section{Conclusion and Discussion}
In this work, we investigated the application of the gamma distribution to model the stationary distributions of populations governed by a discrete stochastic logistic and Ricker equations at steady state. We incorporated power nonlinearities and examining the behavior of the model under various parameterizations. This generalization broadens the applicability of the stochastic logistic framework to more complex ecological scenarios. We established relationships between the parameters of the gamma distribution and the intrinsic growth rate  $r$ of the stochastic equations at equilibrium.  We identified and demonstrated two branches of the intrinsic growth rate, \( r_+ \) and \( r_- \), for \eqref{eq:logistic3} and \eqref{ricker1}. These branches represent alternative stable states, corresponding to higher and lower growth rates, respectively. This duality provides deeper insights into population stability and resilience under stochastic conditions. We explored the ecological implications of these mathematical relationships, highlighting, for instance, that at equilibrium,  $r$ is independent of \(\theta \),  the scale parameter of the gamma distribution. This finding underscores the role of internal biological mechanisms, as reflected by the shape parameter \( k \) in driving population dynamics.

These mathematical relations underscore the critical interplay between  $k$ , $r$, and $Var(\epsilon_t)$, revealing how environmental variability and internal population structure shape dynamic outcomes. The results highlight the sensitivity of population dynamics to environmental variability. Populations with larger \( k \), corresponding to more symmetrical and stable distributions, are more resilient to stochastic effects. Conversely, populations with smaller \( k \) are more prone to instability, particularly when subjected to high variance. Understanding the interplay between \( k \), \( r \), and \( \text{Var}(\epsilon_t) \) can inform conservation strategies by identifying populations at greater risk of extinction. Management practices aimed at reducing environmental variability could stabilize populations and extend the feasible range of \( k \) for stable solutions.

Future research can continue on addressing the challenges identified in this study while expanding its scope. For example, one may analyze more sophisticated discrete stochastic models that incorporate biological complexities such as age structure, migration, or varying environmental conditions. These enhanced models may require the development of innovative mathematical methods to better reflect the properties of population dynamics in real-world scenarios. Moreover, establishing a robust mathematical foundation to rigorously prove the relationships between model parameters and their stationary distributions is crucial. Such work would deepen our understanding of stochastic population behavior and determine when the gamma distribution, or other distributions, serves as a valid and practical approximation for equilibrium states.

Finally, future research could leverage real-world ecological population data to validate the theoretical relationships between stochastic models and the gamma distribution. This validation would involve fitting the stochastic models to empirical data to estimate their parameters, as well as deriving the corresponding gamma distribution parameters. By comparing these empirical estimates with the theoretical predictions derived in this study, researchers can assess the accuracy and applicability of the models under real-world conditions. This process would not only verify the theoretical results but also refine the models to better capture the complexities of ecological systems. Furthermore, validation efforts could explore how well the gamma distribution approximates population dynamics across different environmental contexts, such as varying degrees of stochasticity or specific biological factors like age structure or migration. Such an approach would provide a deeper understanding of how stochastic processes influence population behavior and ensure that the models are robust and applicable to diverse ecological scenarios.

\section{Appendix}
In this section, we prove the identity for $s \geq 0$ 

\begin{equation}
E\left[ X_t^n e^{- s X_t} \right] = \frac{\Gamma(k + n)}{\Gamma(k)} \frac{\theta^n}{(1 + s \theta)^{k + n}}
\label{identity}
\end{equation}
here \( X_t \) be a gamma-distributed random variable with shape parameter \( k > 0 \) and scale parameter \( \theta > 0 \) and its probability density function (PDF) of \( X_t \) is:
\begin{equation}
f_{X_t}(x) = \frac{1}{\Gamma(k) \theta^{k}} x^{k - 1} e^{- x / \theta}, \quad x > 0
\end{equation}

We aim to compute the expected value:

\begin{equation}
E\left[ X_t^n e^{- s X_t} \right] = \int_{0}^{\infty} x^{n} e^{- s x} f_{X_t}(x) \, dx
\end{equation}
Substituting the PDF into the Expected Value gives 

\begin{align}
E\left[ X_t^n e^{- s X_t} \right] &= \int_{0}^{\infty} x^{n} e^{- s x} \left( \frac{1}{\Gamma(k) \theta^{k}} x^{k - 1} e^{- x / \theta} \right) dx \\
&= \frac{1}{\Gamma(k) \theta^{k}} \int_{0}^{\infty} x^{n + k - 1} e^{- x (s + 1 / \theta)} dx
\end{align}
Recall the standard gamma integral

\begin{equation}
\int_{0}^{\infty} x^{\alpha - 1} e^{- \lambda x} dx = \frac{\Gamma(\alpha)}{\lambda^{\alpha}}, \quad \text{for } \lambda > 0, \ \alpha > 0
\end{equation}
Therefore, the integral evaluates to:
\begin{equation}
\int_{0}^{\infty} x^{n + k - 1} e^{- x (s + 1 / \theta)} dx = \frac{\Gamma(n + k)}{\left( s + \dfrac{1}{\theta} \right)^{n + k}}
\end{equation}
and
\begin{align}
E\left[ X_t^n e^{- s X_t} \right] &= \frac{1}{\Gamma(k) \theta^{k}} \cdot \frac{\Gamma(n + k)}{\left( s + \dfrac{1}{\theta} \right)^{n + k}} \\
&= \frac{\Gamma(n + k)}{\Gamma(k)} \cdot \frac{1}{\theta^{k}} \cdot \frac{1}{\left( s + \dfrac{1}{\theta} \right)^{n + k}}
\end{align}
Thus, we have:

\begin{equation}
E\left[ X_t^n e^{- s X_t} \right] = \frac{ \Gamma(k + n) }{ \Gamma(k) } \frac{ \theta^{n} }{ (1 + s \theta )^{ k + n } }
\end{equation}

This completes the verification of the identity.

\end{document}